\documentclass[aerospace,communication,accept,pdftex,moreauthors]{Definitions/mdpi}

\firstpage{1}
\pubvolume{1}
\issuenum{1}
\articlenumber{0}
\pubyear{2026}
\copyrightyear{2026}
\externaleditor{   } 
\datereceived{ }
\daterevised{ } 
\dateaccepted{ }
\datepublished{ }

\Title{A First-Order Assessment of Permanent Magnet Deflection for Space Radiation Protection}

\Author{{Valerio} 
 Parisi $^{1}$\orcidA{}, Roberto Capuzzo Dolcetta $^{1,2,3,4,}$*\orcidB{}, Fabrizio Frezza $^{5}$\orcidC{} and Luca Lunati $^{6}$\orcidD{}}

\AuthorNames{Valerio Parisi, Roberto Capuzzo Dolcetta, Fabrizio Frezza and Luca Lunati}

\address{$^{1}$ \quad Department of Physics, {Sapienza University of Roma}
, 2, {P}
iazzale A. Moro, 00185 Roma, Italy; {valerio.parisi@fondazione.uniroma1.it} 
            \\
$^{2}$ \quad Centro Ricerche Enrico Fermi, 89 A, via Panisperna, 00184 Roma, Italy \\
$^{3}$ \quad Istituto Nazionale di Fisica Nucleare, 2, Piazzale A. Moro, 00185 Roma, Italy \\
$^{4}$ \quad Istituto Nazionale di Astrofisica, 8, {V}iale del Parco Mellini, 00136 Roma, Italy \\
$^{5}$ \quad {Department of Information Engineering, Electronics and Telecommunications 
, Sapienza University of Roma}, 18, via Eudossiana, 00184 Roma, Italy; {fabrizio.frezza@uniroma1.it} \\
$^{6}$ \quad
{GSI Helmholtz-Zentrum, Biophysics Department},  
 1, Planckstraße 1, 64291 Darmstadt, Germany; {l.lunati@gsi.de}
 }

\corres{Correspondence: {roberto.capuzzodolcetta@fondazione.uniroma1.it} 
}

\abstract{We present a preliminary feasibility assessment of a magnetic shield designed to protect a space probe from cosmic radiation via magnetic deflection using neodymium permanent magnets.
This work is grounded in theoretical considerations whose preliminary indications are intended to serve as the basis for future Monte Carlo simulations and laboratory validation.
The novelty of our approach lies in the use of a magnetic shield; its competitiveness with conventional passive absorbing shielding is not investigated here but warrants dedicated future work.
The primary objective is to protect a spacecraft from the flux of charged particles emitted by the Sun. To this end, we combine theoretical modeling and numerical simulations, followed by the construction of a prototype for laboratory testing and, potentially, for future experimental validation at the CubeSat scale.}


\begin{document}



\section{Introduction}\label{sec1}

Space missions beyond low Earth orbit expose astronauts to a harsh radiation environment dominated by Solar Particle Events (SPEs) and Galactic Cosmic Rays (GCRs)~\citep{durante2011physical}. SPEs consist predominantly of protons with energies up to several GeV and can deliver life-threatening doses under unshielded conditions~\citep{kim2009prediction}, whereas GCRs, though less intense, are highly penetrating and responsible for long-term health risks~\citep{cucinotta2000model}. Current radiation protection strategies rely primarily on passive shielding (e.g., polyethylene, aluminum), but the required mass grows rapidly with the desired protection level, making such approaches costly for deep-space missions~\citep{narici2017performances}.

Active shielding using electromagnetic fields has been proposed as a lighter alternative~\citep{townsend2001overview}. However, most concepts employ superconducting magnets, which face significant challenges in spaceflight owing to cryogenic requirements and structural complexity~\citep{spillantini2007shielding}. Permanent magnets, by contrast, offer intrinsic field stability and simplicity; yet, their application to radiation deflection has received little attention.

In this work, we present a first-order feasibility study of a magnetic shield based on neodymium permanent magnets. We develop a simplified analytical model to estimate the deflection of a collimated proton beam (representing SPEs) by a constant magnetic field of finite volume. We provide rough estimates of mass, size, and deflection efficiency for different magnet configurations and compare the performance with conventional passive shielding. The aim is to assess whether permanent magnets can offer a mass-competitive solution for directional protection during solar particle events.

The paper is organized as follows: Section~\ref{sec2} addresses the space radiation environment and the rationale for dedicated protection. Section~\ref{sec3} introduces our concept of radiation protection via a passive magnetic deflector comprising an array of permanent (neodymium) magnets. Section~\ref{sec4} provides preliminary indications of the magnetic shield's efficiency. Finally, Section~\ref{sec5} presents our conclusions and outlines directions for future work.


\section{Space Radiation Protection}\label{sec2}

Conventional shielding relies on passive materials with low atomic number (e.g., polyethylene, water, aluminum) to absorb or fragment incoming particles~\citep{narici2017performances}. However, the required mass scales with stopping power; for a Mars mission, cumulative exposure can reach \(\sim 1\) Sv, exceeding NASA's career limits~\citep{nar26}. A storm shelter providing adequate shielding against a large SPE may weigh up to tens of tons.

Active shielding using electromagnetic fields offers the potential for mass reduction by deflecting charged particles rather than absorbing them. Several concepts have been proposed, including electrostatic shields, plasma shields, and magnetic fields generated by superconducting coils~\citep{townsend2001overview}. Superconducting magnets can produce strong fields (\(\sim\)1 T) over large volumes, but they require cryogenic cooling, which poses reliability and power challenges for spaceflight~\citep{spillantini2007shielding,Battiston12}. Permanent magnets, in contrast, are simple, robust, and require no power, but they produce weaker fields (typically \(0.1\)–\(1\) T at the surface) over limited volumes. Their feasibility for space radiation protection has not been systematically explored. In this work, we assess whether a practical magnetic shield using permanent neodymium magnets could provide a mass-competitive alternative for directional \mbox{SPE protection.}

\section{A Possible Magnetic Shield}\label{sec3}
\begin{figure}[H]
    \includegraphics[width=\textwidth]{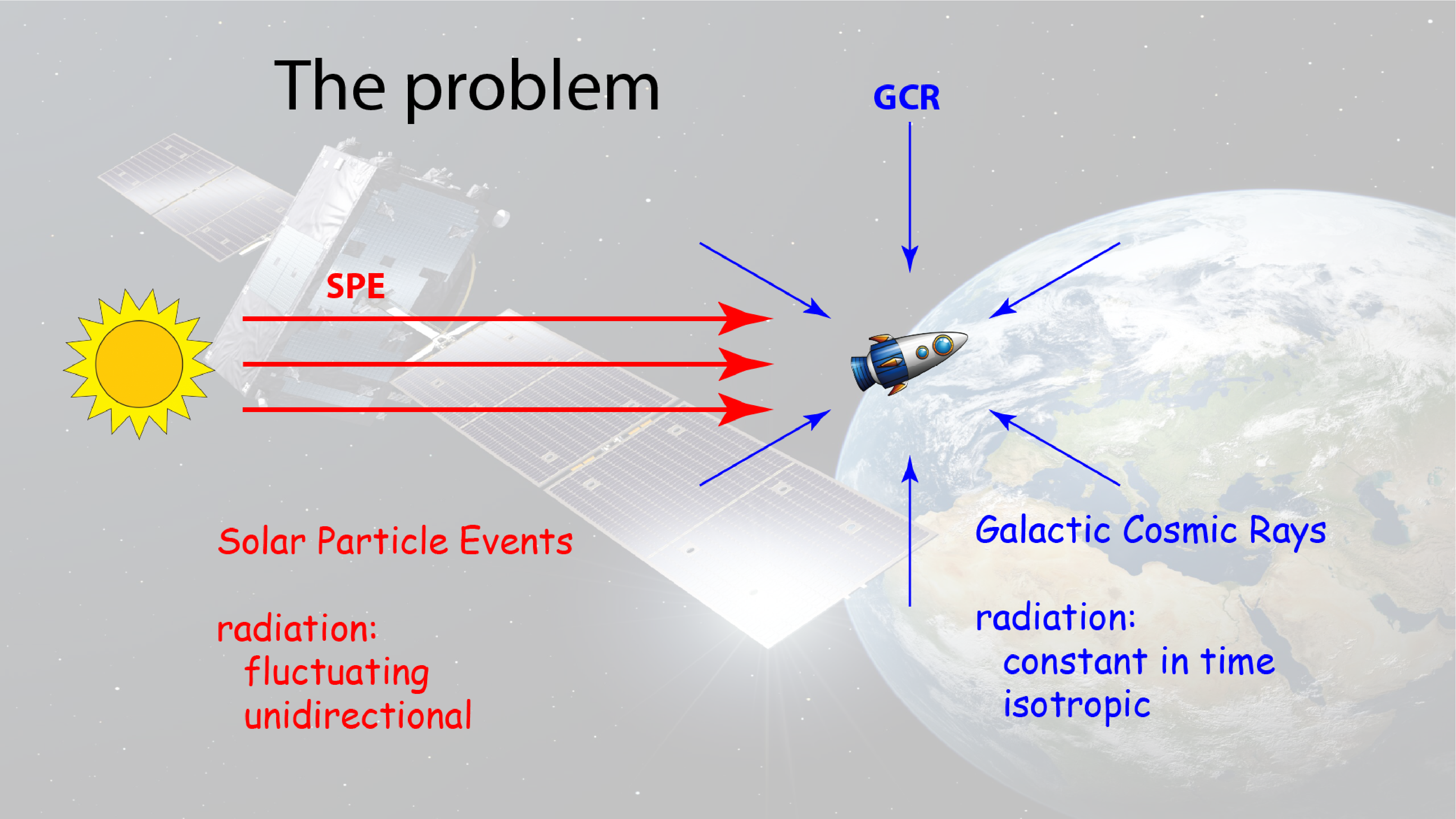}
     \caption{{Simple} 
 schematization of the cosmic bombardment onto a space probe (from \cite{paretal24}).}
    \label{fig:1}
\end{figure}

In light of the preceding considerations, we decided to investigate a radiation shielding concept based on the magnetic deflection of incoming charged particles, aiming to provide protection while keeping the shield mass within practical limits for space missions~\citep{paretal24}. Figure \ref{fig:1} provides a conceptual sketch of the problem.
This is, in principle, achievable using permanent magnets, but it will, of course, require a thorough analysis of efficiency, feasibility, and cost.

In this paper, we address the theoretical foundations of our proposed shielding mechanism, providing preliminary indications of its feasibility.

\subsection*{\highlighting{Magnetic} 
 Deflection of a Collimated Charged-Particle Flux}
As an initial step, we approach the problem using several simplifying approximations.

The effectiveness of a permanent-magnet system for deflecting collimated particle radiation can be estimated using a relatively simple approach. This approach is intended to determine whether further development of the shielding system is warranted.
It is worth recalling that our scheme, as proposed below, aims at the efficient deflection of particles arriving from a given direction and within a specific energy range. This scenario applies well to active Sun conditions, when bursts of high-energy protons are emitted and may pose a hazard to a space probe along its trajectory.

The next phase of the study will involve extensive practical simulations, both numerical and laboratory-based, followed by testing on specific space missions scaled to the CubeSat level.
In this way, we will properly assess the actual capability of the proposed mechanism in terms of reliability, efficiency, and estimated costs, with a view to possible practical application in crewed space missions.

The framework adopted is the study of the performance of a single cylindrical permanent magnet in deflecting a hypothetical, collimated (i.e., non-isotropic) beam of identical and monoenergetic charged particles. This type of beam serves as an idealized representation of SPEs.

The additional assumption of a simplified estimate of the magnetic field produced by the magnet in its surrounding environment will provide
a reliable estimation of the system's deflection effectiveness through simple geometric evaluations, without the need for numerical simulations.

\begin{figure}[H]
\includegraphics[width=0.8\linewidth]{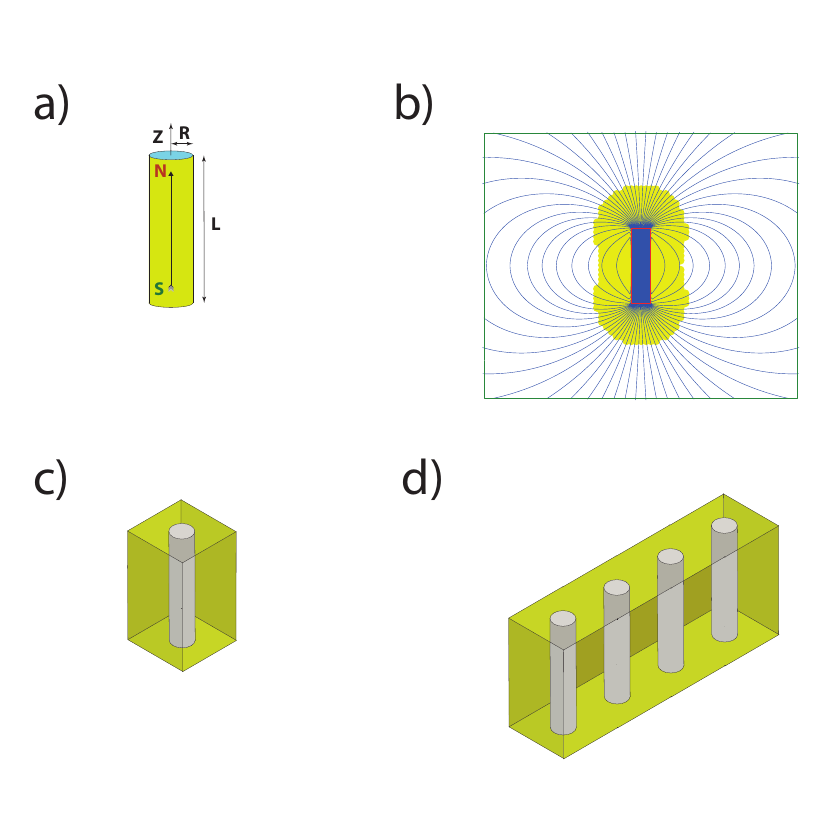}
\caption{{Panel} 
 {(\textbf{a}):} 
 single magnetic cylinder; panel (\textbf{b}): field lines of the single {cylindrical magnet (in blue, while in yellow the region of almost constant field)}; panel (\textbf{c}): the magnetic field generated by the cylindrical permanent magnet (in grey) is approximately constant within the (yellow) parallelepiped and nearly zero outside; panel (\textbf{d}):
array of four cylindrical permanent magnets producing an approximately constant magnetic field in the yellow volume and nearly zero outside.
}
\label{fig:2}
\end{figure}

Let us consider a cylindrical permanent magnet of length $L$ and radius $R$, with symmetry axis $z$ such that $z=0$ on the upper surface (see Figure \ref{fig:2}). The magnetic flux density along $z$ is given {by} 

\begin{equation}
B(z)=\frac{B_0}{2}\left(\frac{L+z}{\sqrt{R^2+(L+z)^2}}-\frac{z}{\sqrt{R^2+z^2}}\right).
\label{eq:1}
\end{equation}

From Equation \eqref{eq:1}, if $L\gg R$, at the center of the magnet ($z=-L/2$) we have $B\approx B_0$,
while on both the top ($z=0$) and bottom ($z=-L$) plane surfaces of the magnet, $B\approx B_0/2$.

In the case $L\gg R$, within a volume of order $L^3$ containing the magnet, the magnetic field intensity remains of the same order of magnitude.
This implies that, as a first-order approximation for evaluating the deflection capability of the cylindrical magnet, we may consider the magnetic field generated by the cylindrical magnet as approximately constant within a sphere of radius $(\sqrt{2}/2)L$ containing the magnet, and zero outside it.
For a cylindrical (permanent) neodymium magnet characterized by a fairly standard value $B_0=1$ T, with a specific weight of $\rho =7$ g cm$^{-3}$~\citep{Coey2002, Coey2020, HeimVanderWal2023, MaltsevaEtAl2022}, we considered three cases:

\begin{itemize}
\item $R=10$ cm and $L=100$ cm; it weighs \(\sim 56\) kg and produces a field of order \(10^{-2}\) T over a volume of a few m\(^3\);
\item $R=1$ cm and $L=100$ cm; it weighs \(\sim 0.6\) kg, which is about two orders of magnitude less than the previous case, producing a field also about two orders of magnitude weaker (\(\sim\)\(10^{-4}\) T) over the same volume of a few m\(^3\);
\item $R=10$ cm and $L=1000$ cm; it weighs \(\sim 560\) kg and produces a field of order \(10^{-4}\) T over a volume of a few thousand m\(^3\);
\end{itemize}

\vspace{-30pt} 

\begin{figure}[H]
   \centering
    \includegraphics[width=0.4\textwidth]{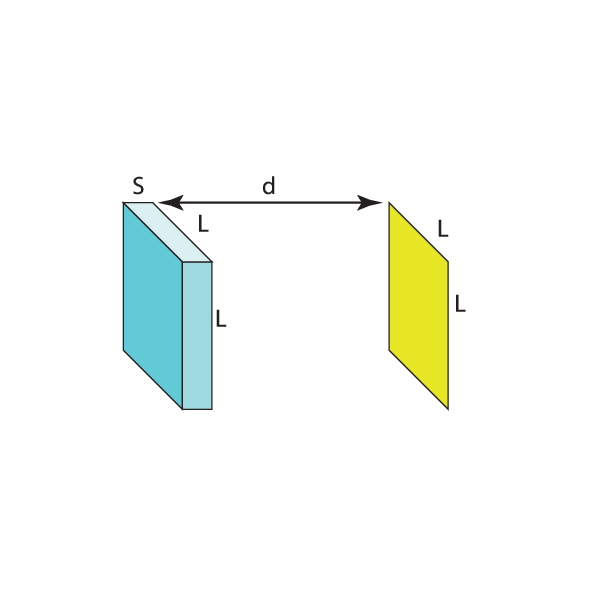}\\[-10ex]
    \includegraphics[width=0.4\textwidth]{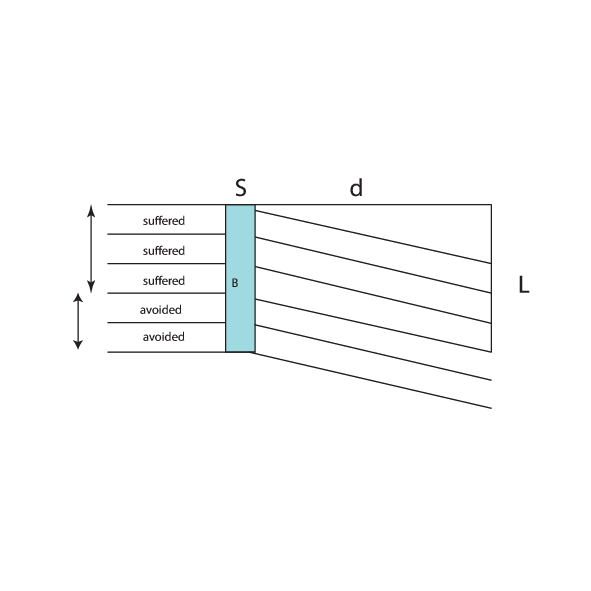}\\[-10ex]
    \includegraphics[width=0.4\textwidth]{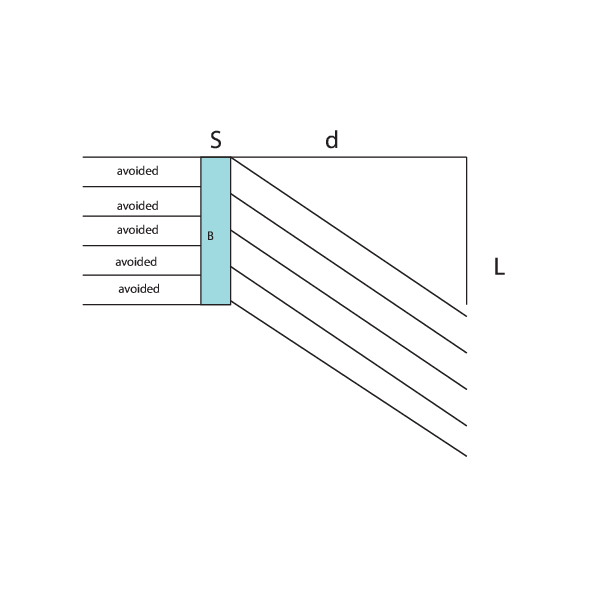}\\[-25ex]
    \includegraphics[width=0.8\textwidth]{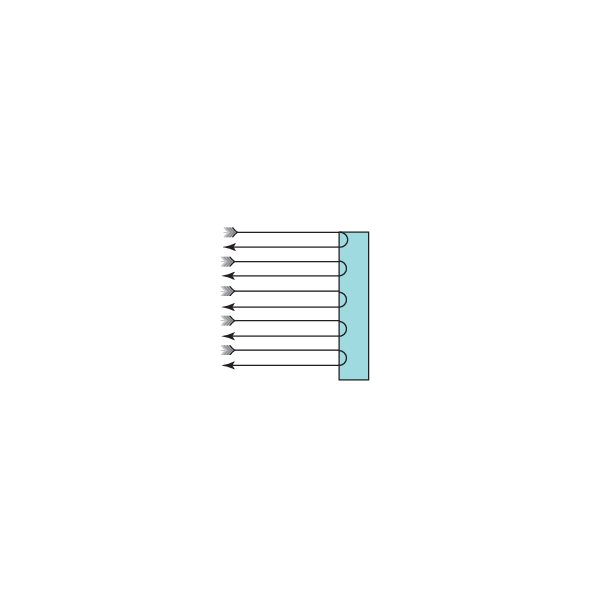}\\[-20ex]

    \caption{{Deflection} 
 {schemes,} 
 {ranging top-down} 
 from partial to total protection {(lowest panel)} of the $L\times L$ target by a magnetic $L\times L \times S$ shield placed at distance $d$.}
 
    \label{fig:3}
\end{figure}

With reference to Figures \ref{fig:2} and \ref{fig:3}, consider a magnetic field of intensity $B=\text{const.}$ inside a parallelepiped of dimensions $L\times L \times S$ and $B=0$ outside. To completely deflect unidirectional and monochromatic particle radiation from a target of size $L \times L$, the deflection angle $\theta$ between the incoming particle trajectory and the deflected trajectory must satisfy $\theta> \arctan (L/d)$, where $d$ is the distance between the parallelepipedal screen and the target.

For incoming protons of $1$ MeV kinetic energy, the gyroradius in the presence of a constant $B=10^{-2}$ T is $14.5$ m. Taking $L=10$ m and $S=5$ m, the proton trajectory deflection angle should be at least
$\theta = \arcsin{\left({5}/{14.5}\right)} \simeq 20.2^{\circ}$. Consequently, the distance between the screen and the target must be $d > L/\tan{\theta} = 10 \textrm{m}/\tan{20.2^{\circ}} = 27.2$ m.

Of course, this threshold distance increases with a reduction in the magnetic field strength and/or an increase in the incoming proton energy.

Preliminary estimates indicate that the goal of efficiently deflecting (20\%) SPEs at moderate energies (0.1~MeV $\leq E \leq$ 10~MeV) could be achieved while keeping the weight below 300~kg by employing an array of 1482 Neodymium–Iron–Boron (NdFeB) cubic magnets (the most powerful commercially available), each of size 3~cm \(\times\) 3~cm \(\times\) 3~cm, arranged on an approximately square ($1.17 \times 1.14$~m) 2D grid. The grid dimensions correspond to 39 magnets along the $1.17$~m side and 38 along the $1.14$~m side, yielding $39 \times 38 = 1482$ magnets. Each magnet provides a surface field of $\sim$1~T. This structure is expected to deflect about $20\%$ of the incoming solar particles, thereby protecting a target of area comparable to the battery shield at a distance of $2$--$3$~m.

On the other hand, a configuration based on cylindrical magnets instead of cubic ones, with similar performance but stronger directionality in the generated field, would comprise 1283 magnets of 3~cm diameter and 3~cm height, packed in a honeycomb pattern over a comparable area. The number 1283 is derived from the hexagonal close‑packing density ($\pi/(2\sqrt{3}) \approx 0.9069$) applied to the same $1.17 \times 1.14$~m area, accounting for the footprint of each cylinder. The weight would be approximately 230 to 280~kg, with a high degree of modularity. We believe this weight could be competitive with traditional passive shielding, although this merits further investigation. The advantage of cubic magnets is that they completely fill the container structure without leaving voids. Conversely, cylindrical magnets offer better directionality, i.e., a more focused field than cubic magnets, and are therefore better suited for deflecting charged particles arriving from a known direction.

\section{Preliminary Indications on the Magnetic Shield Efficiency}\label{sec4}

\begin{figure}[H]
    \includegraphics[width=1.0\linewidth]{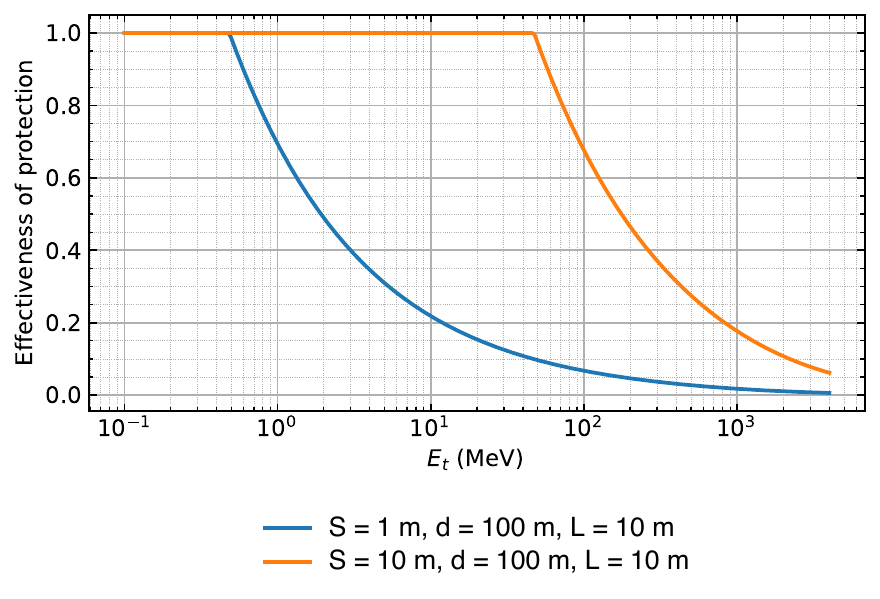}
    \caption{Protection effectiveness as a function of the incoming proton energy for $B=0.01$ T, for the two example cases with parameter sets $(S, d, L)$.}
    \label{fig:4}
\end{figure}

In the context of developing a suitable architecture for a magnetic radiation shield, it is necessary to obtain preliminary indications of both the efficiency and feasibility of such devices.
To this end, Figure \ref{fig:4} shows the
protection \textit{{effectiveness}
} of a device of the type described in the previous section for a field intensity of $B=0.01$ T, and for two example cases: ($S=1$ m, $d=100$ m, $L=10$ m) and ($S=10$ m, $d=100$ m, $L=10$ m). Effectiveness is defined as the fraction of solar protons actually deflected from their impact trajectory toward the target.

\begin{figure}[H]
    \includegraphics[width=1.0\linewidth]{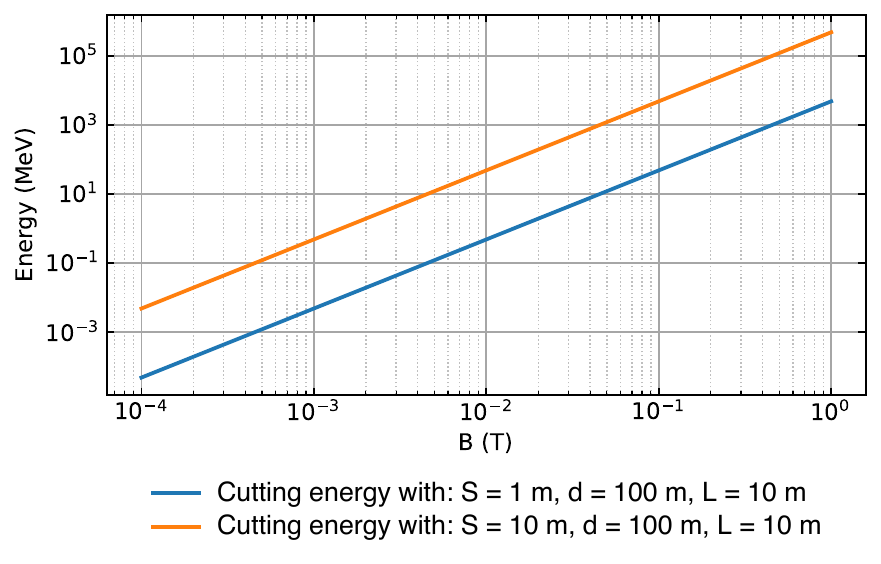}

    \caption{Cutting (threshold) energy as a function of $B$ for two cases.}
    \label{fig:5}
\end{figure}

Note that the device essentially acts as a high-pass filter, deflecting all protons with energies below an energy threshold \(E_t\), which, in the examples of Figure \ref{fig:4}, is \(E_t=0.5\) MeV and \(E_t= 50\) MeV, while maintaining \(50\%\) protection up to \(\sim\)\(6E_t\). The most important parameter is therefore the threshold energy \(E_t\), which is shown as a function of the magnetic field intensity in Figure \ref{fig:5} for the same example cases of Figure \ref{fig:4}.

Regarding the energy distribution of incoming solar protons, we refer to the comprehensive catalog of solar energetic protons compiled from SOHO/ERNE data for solar cycles 23 and 24~\citep{mit24}. It appears that most SPEs fall within the approximate range \mbox{20--100 MeV.}

\begin{figure}[H]
    \includegraphics[width=1.0\linewidth]{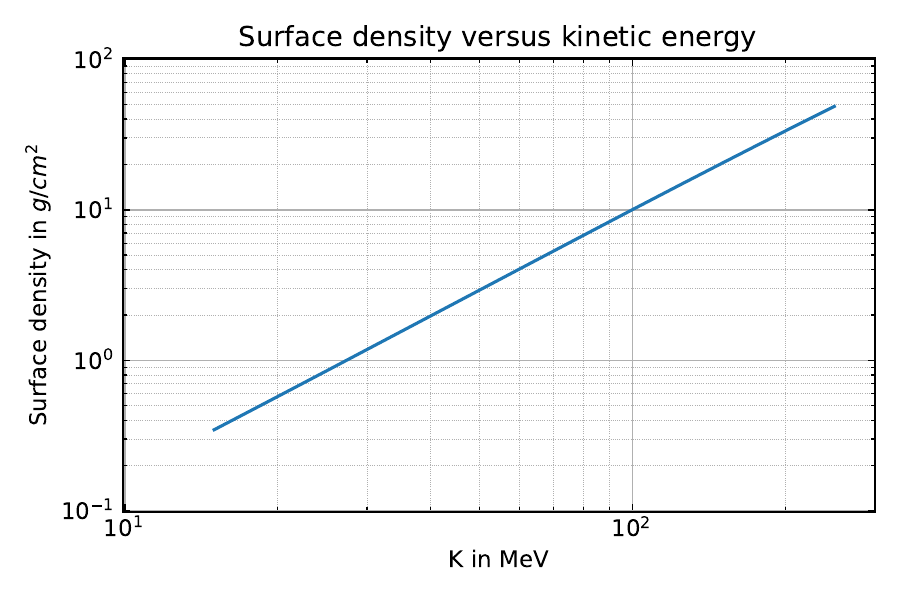}
    \caption{{Surface} 
 density (g cm$^{-2}$) of an aluminum shield required to stop mono-energetic protons of kinetic energy K, as computed from NIST PSTAR CSDA range data
\citep{ber99}.}
    \label{fig:6}
\end{figure}

Finally, in Figure \ref{fig:6}, we plot, as a useful comparison, our evaluation of the surface density of an aluminum protection shield required to screen against mono-energetic protons of a given kinetic energy.

\vspace{-15pt} 

A precise comparison between the performance and potential advantages of the magnetic screen and the standard passive screen would require an analysis of numerous aspects and degrees of freedom in the realization of the magnetic shield (geometry of the magnet distribution, granularity of the magnetic structure, dimensions and location of the structure, etc.). This lies beyond the scope of the present paper but is certainly of great importance for the eventual realization of the device.

Here, we provide only rough indications. The 1/100 Tesla field, which would protect against protons of energies up to a few MeV and a few tens of MeV in the cases of the upper and lower curves of Figure \ref{fig:5}, respectively, can be produced, for example, within a \mbox{1 m-size} cubic distribution of neodymium magnets corresponding to a volume density of 50 kg/m\(^3\)---which indeed corresponds to a screen weight of 50 kg.

Finally, two aspects should be considered when comparing magnetic and material shielding. First, below a certain energy threshold, a magnetic field can provide complete protection by deflecting charged particles away from the protected volume, whereas aluminum shielding only reduces the particle flux through energy loss and attenuation. Second, in this paper, the magnetic field produced by the permanent magnets has been approximated as uniform and confined to a cubic region; this simplification is conservative, because real magnetic field lines extend outside the cube and may further enhance particle deflection and, consequently, overall shielding effectiveness.

\section{Conclusions}\label{sec5}

This work presents a preliminary assessment of a magnetic shielding strategy for space radiation protection, based on the active deflection of charged particles via permanent neodymium magnets.
Our primary motivation is the critical need to mitigate exposure to high-energy particles, particularly during Solar Particle Events. To this end, we have analyzed the viability of generating localized magnetic fields capable of altering the trajectories of incident charged particles.
We  note, however, that large SPE storms may be accompanied by bursts of electromagnetic radiation that could also cause damage. For such events, the passive magnetic shield is, of course, ineffective.

A possible extension of our proposed approach to the persistent flux of Galactic cosmic rays lies beyond the scope of this paper and is left for future investigation.


The adoption of an active magnetic shield would certainly be advantageous in terms of weight compared with complete passive shielding of a crewed spacecraft, whose surface area cannot be less than approximately $100$ m\(^2\).

The simplified model presented here considers the magnetic field generated by cylindrical permanent magnets to provide an initial estimate of the deflection capability for mono-energetic, collimated particle beams. These approximations, while neglecting complex particle dynamics and magneto-geometrical interactions, offer insight into the scale of field intensities and volumes required for effective deflection. The analysis indicates that magnetic fields of the order of $10^{-2}$ T can produce significant deflection angles for low-energy protons (e.g., 1 MeV) over practical distances, suggesting that directional shielding of sensitive spacecraft regions is feasible under realistic constraints.

It is important to note that the magnetic field strength inside the spacecraft would be low enough to be harmless to the crew and not interfere with internal electronic devices (with the possible exception of magnetic compasses or sensors).

The proposed scheme overcomes some of the inherent limitations of superconducting magnet systems, such as cryogenic requirements and mechanical complexity, by exploiting the intrinsic field stability and structural simplicity of permanent magnets. However, this comes at the cost of reduced field tunability and potentially limited shielding volume.

Further development will involve detailed numerical simulations of particle--magnet interactions under realistic space radiation spectra, multi-particle species, and angular distributions, as well as the inclusion of spacecraft structural effects and secondary radiation production. Laboratory validation using charged-particle beams and scaled magnet configurations will serve as a precursor to potential CubeSat-level in-orbit demonstrations.

Ultimately, the results support the plausibility of a magnetostatic shielding architecture as a complementary or alternative strategy to conventional passive systems, particularly for targeted protection during transient high-flux events. This approach could represent a scalable and mass-efficient component within an integrated space radiation protection framework for future crewed missions beyond low Earth orbit.

Finally, although preliminary, this concept represents a potentially scalable element in an integrated hybrid shielding system combining magnetic deflection and lightweight passive materials.

\subsection*{Limitations and Open Issues}

We acknowledge that the present first-order assessment rests on several simplifying assumptions that will need to be addressed in subsequent work. First, our analytical model assumes a perfectly collimated, monoenergetic proton beam and a constant magnetic field over a finite volume; realistic SPE spectra, angular distributions, and field inhomogeneities require numerical simulations (e.g., Geant4 {11.4.1} or COMSOL {6.3}) 
 to validate the deflection estimates. Second, the interaction of primary protons with the high‑Z materials (Nd, Fe, B) of the permanent magnets may generate secondary neutrons and gamma rays, whose additional dose contribution should be evaluated. Third, long‑term exposure to space radiation could partially demagnetize the neodymium magnets, potentially reducing their effectiveness over multi‑year missions; experimental studies on demagnetization under simulated space conditions are needed. Results in this regard are necessarily very preliminary and difficult to apply to realistic long-duration space travel scenarios (for a discussion of this topic, see \cite{sam18}). Fourth, although the expected static magnetic field inside the spacecraft is low ($\leq$0.01 T) and generally considered safe for crew and most electronics, its electromagnetic compatibility with sensitive on‑board instruments (e.g., magnetometers, compasses) requires case‑by‑case verification. Finally, a full trade‑off analysis comparing the mass, volume, and cost of the proposed magnetic shield with optimized passive shielding (e.g., polyethylene or water) is left for a dedicated follow‑up study. Addressing these aspects will form the core of our future research, including Monte Carlo simulations, laboratory beam tests, and a potential CubeSat‑based in‑orbit demonstration.






\vspace{6pt}

\authorcontributions{Conceptualization, V.P., R.C.D., F.F. and L.L.; methodology, R.C.D., V.P., F.F. and L.L.; resources, F.F.; data curation, R.C.D. and V.P.; writing---original draft preparation, R.C.D.; writing---review and editing, R.C.D. and V.P.; project administration, F.F.; funding acquisition, F.F. and V.P. All authors have read and agreed to the published version of the manuscript.}

\funding{This work was supported by the SPACE IT UP{!} (Spoke 9) Italian funding.}

\dataavailability{No new data were created or analyzed in this study. Data sharing is {not applicable} 
 to this article.}

\acknowledgments{We warmly acknowledge {Giuseppe} 
 Presta for discussions on the topics of \mbox{this paper.}}

\conflictsofinterest{The authors declare no conflicts of interest.}


\pagebreak

\begin{adjustwidth}{-\extralength}{0cm}

\reftitle{References}


\PublishersNote{}

\end{adjustwidth}




\end{document}